\documentclass[preprint,review,12pt]{elsarticle}
\usepackage{graphicx}
\usepackage{epsf}
\usepackage{amsmath}
\usepackage{lscape}
\usepackage{color}
\usepackage{dcolumn}
\usepackage{hyperref}
\usepackage{amsmath}
\usepackage{rotating}
\usepackage{soul}
\usepackage{longtable}
\usepackage{natbib}

\newcommand{\3}{$_{3}$}
\newcommand{\4}{$_{4}$}

\newcommand{\ai}{{\it ab initio}}

\newcommand{\ea}{\emph{et al}.}
\newcommand{\etal}{\emph{et al}.}

\newcommand{\mol}[1]{{\bf #1}}

\journal{Molecular Astrophysics}

\begin{document}

\begin{frontmatter}

\title{Laboratory spectra of hot molecules: data needs for hot super-Earth exoplanets}

\author{Jonathan Tennyson and Sergei.~N. Yurchenko}
\address{Department of Physics and Astronomy, University College London,London, WC1E 6BT, UK}

\begin{abstract}

The majority of stars are now thought to support exoplanets. Many of those
exoplanets discovered thus far are categorized as  rocky
objects with an atmosphere. Most of these objects are however hot  due to their
short orbital period. Models suggest that water is the dominant species in their atmospheres.
The hot temperatures are expected to turn these
atmospheres into a (high pressure) steam bath containing remains of melted rock.
The spectroscopy of these hot rocky objects  will be very different from that of
cooler objects or  hot gas giants.
Molecules
suggested to be important for the spectroscopy of these objects are reviewed
together with the current status of the corresponding spectroscopic data.
Perspectives of building a comprehensive database of linelist/cross sections applicable for
atmospheric models of rocky super-Earths as part of the ExoMol project are discussed. The
quantum-mechanical approaches used in linelist productions  and their challenges
are summarized.

\end{abstract}

\end{frontmatter}

\section{Introduction}

There are vast areas of the Universe  thinly populated by
molecules which are cold. However, there are also huge numbers of
important astronomical bodies which support hot or highly-excited
molecules.  It is the spectroscopic demands of studying these hot
regimes we focus on in this review. We will pay particular attention
to the demands on laboratory spectroscopy of a recently identified
class of exoplanets known as hot rocky super-Earths or, more
colourfully, lava and magma planets. These planets orbit so close to
their host stars that they have apparent temperatures such that their
rocky surface should melt or even vaporize. Little is known about
these planets at present: much of the information discussed below
is derived from models rather than observation.

Of course hot and cold are relative terms; here we will take room
temperature ($T \sim 300$ K)
as the norm which means, for example, that so-called cool
stars which typically have temperatures in the 2000 -- 4000 K range
are definitely hot.  Much of the cold interstellar medium is not
thermalised and excitation, for example by energetic photons, can lead
to highly excited molecules.  This can be seen, for example, from maser
emissions involving transitions between highly
excited states, which is observed from a range of molecules from a variety of
interstellar environments \cite{12Gray.book}.
Similarly the coma of comets are
inherently cold but when bathed in sunlight can be observed to emit
from very high-lying energy levels \cite{jt330,jt349,jt452}.

Turning to the consideration of exoplanets.  At the present it even
remains unclear how to conclusively identify which planets of a few to
ten Earth masses are actually rocky \cite{17TaTaHe}.  From density
observations some of them appear to be rocky (silicate-rich), or with
a fraction of ice/iron in the interior.  Others suggest a structure
and composition more similar to gas giants like Neptune.  Density
alone is not a reliable parameter to distinguish among the various
cases.  In addition to there is a class of ultra-short period (USP)
exoplanets which are thought to be undergoing extreme evaporation of
their atmosphers due to their close proximity to their host star
\cite{12GiTrFo,14SaRaWi,14GiAnCo,17ObRoCo}. These objects
are undoubtedly hot but as yet there are  no 
mass measurements for USP planets.  Spectroscopic
investigations of atmospheres of super-earths and related
exoplanets holds out the best prospect of
learning about these alien worlds.  The prospects of observing the
atmospheric composition for the transiting planets around bright stars
make us confident we will be in a much better position in a few years
time with the launch of the James Webb space telescope (JWST) and
future dedicated exoplanet-characterization missions.

From the laboratory perspective, the observation of hot or highly
excited molecules places immense demands on the spectroscopic data
required to model or interpret these species. As discussed below, a
comprehensive list of spectroscopic transitions, a line list, for a single
molecule can contain significantly more than $10^{10}$ lines.  This
volume of data points to theory as the main source of these line
lists \cite{jt511}.

A line list consists of an extensive list of transition frequencies
and transition probabilities, usually augmented by other properties
such as lower state energies, degeneracy factors and partition
functions to give the temperature dependence of the line and, ideally,
pressure-broadening parameters to give the line shape.  For radiative
transport models of the atmospheres of hot bodies, completeness of the
the line list to give the opacity of the species is more important
than high (``spectroscopic'') accuracy for individual line positions.
This is also true for retrievals of molecular abundances in exoplanets
based on the use of transit spectroscopy which, thus far, has largely
been performed using observations with fairly low resolving power ($R
< 3000$)
\cite{13TiEnCo.exo}.  However, the situation is rather different with
the high-dispersion spectroscopy developed by Snellen and
co-workers~\cite{13DeBrSn.exo,13BiDeBr.exo,14BrDeBi.exo,14Snellen,17BiDeBr},
which is complementary to transit spectroscopy.  This technique tracks
the Doppler shifts of a large number of spectroscopic lines of a given
species, by cross-correlating them to the reference lab data on the
line positions. This exciting but challenging technique requires
precise frequencies with $R \geq$ 100,000, as well a good
spectroscopic coverage (hot transitions), available laboratory data is
not always precise enough for this technique to work
\cite{15HoDeSn.TiO}.

This review is organised as follows. First we summarise what is known
about hot rocky super-Earth exoplanets. We then consider the laboratory
techniques being used to provide spectroscopic data to probe the
atmospheres of these bodies and others with similar temperatures. In
the following section we summarise the spectroscopic data available
making recommendations for the best line list to use for studies
of hot bodies. Molecules for which little data appears to be available
are identified. Finally we consider other issues associated with
spectroscopic characterization of lava planets and prospects for the future.

\section{Hot rocky super-Earths}


As of the end of 2016 there are  well over  100 detected exoplanets which
are classified as hot super-Earths. These planets are ones which are considered
to be rocky, that is with terrestrial-like masses and/or radii, see
e.g. Seager {\it et al.} \cite{07SeKuHi}, and which are hot enough for, at least on their
dayside, their rock to melt \cite{16KiFeSc}.  Only a handful of these
planets are amenable to spectroscopic characterization with current
techniques \cite{16MaAgMo}, which makes these few objects the ones suitable
for atmospheric follow-up observations.  All these rocky planets have
very short orbits, meaning that they are close to their star
and hence have hot atmospheres ($ T \gg 300$ K). Some of\
these planets are evaporating with water vapor as a major constituent
of the atmosphere
\cite{09LeRoSc.exo,11BaBoBr,13BaBuHo,13BoAgFr,15Marexx,16DaHiPe}. The
atmospheres of these planets are thought to have a lot in common with the young Earth
\cite{74AlArxx} and the atmosphere of a rocky planet immediately after
a major impact planet is expected to be similar \cite{14LuZaMa}.
However, we note that as they are generally tidally-locked to their host star, hot rocky
super-Earths will generally have significant  day-night temperature gradients
\cite{16DeGide}.

According to the NASA Exoplanets Archive
(\url{exoplanetarchive.ipac.caltech.edu}), key hot exoplanets with masses
and radii in the rocky-planet range include CoRoT-7b, Kepler-10b, Kepler-78b,
Kepler-97b, Kepler-99b, Kepler-102b, Kepler-131c, Kepler-406b,
Kepler-406c, and WASP-47e, with Kepler-36b and Kepler-93b being
slightly cooler than 1673 K
\cite{09LeRoSc.exo,11BaBoBr,11HaFrNa,12CaAgCh,13HoSaMa,13MoDeGu,13PeCaLa,14WeMaxx,15DaWiAr}
Most of the rocky exoplanets that have so far been studied are
characterized by the high temperature of their atmospheres, e.g.,
about 1500~K in Kepler-36b and Kepler-93b, 2474 $\pm$ 71 K in CoRoT-7b
\cite{11LeGrFe.exo}, 2360 $\pm$ 300~K in 55 Cnc e  \cite{jt629,12DeGiSe.exo},
and around 3000 K in Kepler-10b \cite{16KiFeSc}. Somewhat cooler but still hot rocky planets include
temperatures of
700~K in Kepler-37b \cite{13BaRoLi}, 750~K in Kepler-62b~\cite{13BoAgFr}, 580~K in Kepler-62c~\cite{13BoAgFr},
and 400--500~K in GJ 1214b \cite{09ChBeIr,10BeKeMi,12HoBuxx}.


If the main constituent of these atmospheres is steam, it
will heat the surface of a planet to (and above) the melting
point of rock  \cite{88ZaKaPo}.  For example, the continental crust of a
rocky super-Earth should melt at about 1200~K \cite{11SaCweBr},
while a bulk silicate Earth at roughly 2000~K \cite{12ScLoFe.exo}.
The gases are released from the rock as it
heats up and melts, including silica and other rock-forming elements, and is then
dissolved in steam \cite{16FeJaWi}.   The main
greenhouse gases in the  atmospheres of hot rocky super-Earths are steam
(from vaporizing water and hydrated minerals) and
carbon dioxide (from vaporizing carbonate rocks), which lead to development of a massive steam
atmosphere closely linked to magma ocean at the planetary surface
\cite{86MaAbxx,88AbMaxx,88Kasting,88ZaKaPo,08ElSexx.exo,13LeMaCh,16FeJaWi}.


At temperatures up to 3000~K, and prior to significant volatile loss,
the atmospheres of rocky super-Earth are thought to be dominated by \mol{H$_2$O}
and \mol{CO$_2$}\footnote{Molecules thought to be important for the spectroscopy of hot super-Earths are given in bold
when first mentioned.} for pressures above $~$1 bar.
\cite{12ScLoFe.exo}. These objects will
necessarily have spectroscopic signatures which differ from those of
cooler planets. At present interpretation of such signature is
severely impacted by the lack of the corresponding spectroscopic data.
For example, recent analysis of the transit spectrum of 55 Cnc e
\cite{jt629} between 1.125 and 1.65 $\mu$m made a tentative detection
of hydrogen cyanide (\mol{HCN}) in the atmosphere but could not rule out the
possibility that this signature is actually in part or fully due to
acetylene (\mol{HCCH}) because of the lack of suitable laboratory data on
the hot spectrum of HCCH.  The massive number of potential absorbers
in the atmosphere of these hot objects also have a direct effect on the
planetary albedo \citep{88Kasting} as well as
the cooling and hence evolution of the young hot objects; comprehensive data is also crucial to model these processes.

Atmospheric retrievals for hot Jupiter exoplanets
such as HD~209458b, GJ~1214b and HD~189733b \cite{16SiFoNi} show that
transit observations can help to establish the bulk composition of a
planet. However, it is only with good predictions of likely
atmospheric composition allied to a comprehensive database of spectral
signatures  and proper radiative transfer treatment  that the observed spectra
 can be deciphered.
The completeness of the opacities plays a special role in such retrievals:
missing or incomplete lab data when analysing transit data will lead to
overestimates of the corresponding absorbing components.

\begin{figure}
\begin{center}
\includegraphics[width=0.9\textwidth]{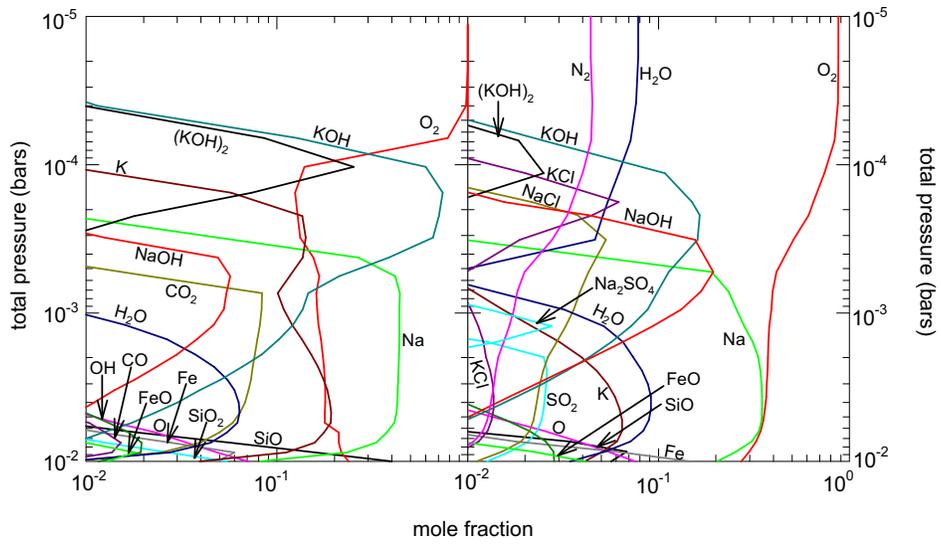}
\caption{Atmospheric composition for a planet similar to CoRoT-7b. Starting compositions
were taken for the continental crust (left) and the bulk silicate Earth  (right) at 2500 K and $10^{-2}$ bars.
Reproduced with
permission from Schaefer {\it et al.}\protect\cite{12ScLoFe.exo}.
\label{f:compos} }
\end{center}
\end{figure}

The typical compositions of steam atmospheres have been considered by
Schaefer {\it et al.}
 \cite{12ScLoFe.exo}, with an example for low atmospheric pressure shown in
Fig.~\ref{f:compos}. The chemical processes on these objects are very
similar to the young Earth (Mars or Venus) and have been studied in
great detail.  The major gases in steam atmospheres (equilibrated chemistry)
with pressures above 1 bar and surface temperatures above 2000 K  are
predicted to be
H$_2$O, CO$_2$, \mol{O$_2$}, \mol{HF}, \mol{SO$_2$}, \mol{HCl}, \mol{OH}, \mol{CO} with continental crust
(CC) magmas (in order of decreasing abundance (mole fractions $~$ 0.8
-- 0.01)) and H$_2$O, CO$_2$, SO$_2$, H$_2$, CO, HF, \mol{H$_2$S}, HCl, \mol{SO},
for bulk silicate Earth (BSE) magmas.  Other gases thought to be present but with smaller mole
fractions ($~$0.01 -- 0.001) include \mol{NaCl}, \mol{NO}, N$_2$, \mol{SO$_3$}, and
\mol{Mg(OH)$_2$} \cite{16FeJaWi}.

\begin{figure}
\begin{center}
\includegraphics[width=.86\textwidth]{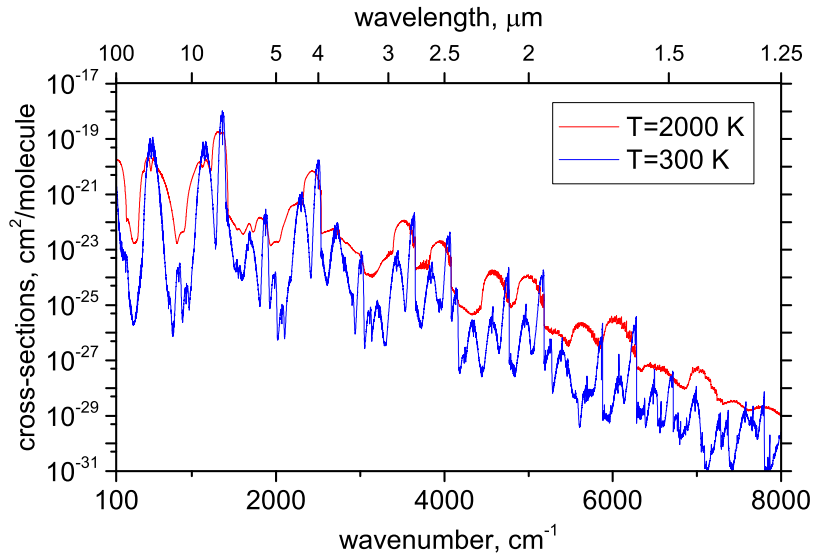}
\caption{Absorption spectrum of SO$_2$ at $T=$ 300~K and 2000~K
simulated using the ExoAmes line list \cite{jt635}.}
\label{f:SO2}
\end{center}
\end{figure}

At temperatures above about 1000~K, sulfur dioxide would enter the atmosphere,
which leads  the exoplanet's atmosphere to be like Venus's, but with steam.
SO$_2$ is a spectroscopically important molecule that is generally
not included in models of terrestrial exoplanet atmospheric models
\cite{12ScLoFe.exo}.
In high concentrations (greater than a few ppm), more than one spectral feature
features
of SO$_2$ are detectable even in low resolution between 4 and 40 $\mu$m
\cite{10KaSaxx}, see also Fig.~\ref{f:SO2}.
This suggests that SO$_2$ should be included when generating models of
atmospheric spectra for terrestrial exoplanets \cite{12ScLoFe.exo}. At high
temperatures and low pressures SO$_2$ dissociates to SO \cite{12ScLoFe.exo}.
Other atmospheric constituents of Venus-like exoplanets include CO$_2$, CO,
SO$_2$, OCS, HCl, HF, H$_2$O, H$_2$S \cite{11ScFexx}.

Kaltenegger {\it et al} \cite{10KaHeSa} studied vulcanism of rocky planets and
estimated the observation time needed for the detection of volcanic activity. The  main sources
of emission were suggested to be H$_2$O, H$_2$, CO$_2$, SO$_2$, and H$_2$S.
Again SO$_2$
should be detectable at abundances of a few ppm for wavelengths between 4 and 40
$\mu$m.


Apart from SO$_2$, significant amounts of \mol{CH$_4$} and \mol{NH$_3$} are expected,
especially  in BSE atmospheres at low temperatures. Although photochemically
unstable, these gases are spectroscopically important and should be considered
in  spectroscopic models of  atmospheres. When sparked by lighting, they combine
to form amino acids, as in the classic Miller-Urey experiment on the origin of
life \cite{59MiUrxx}. Models of exoplanets suggest that NO and \mol{NO$_2$},
as well as a number of other species, are likely to be key products
of lightning in a standard exoplanet atmosphere \cite{jt681}.
Further thermochemical and photochemical processing of the quenched CH$_4$
and NH$_3$ can lead to significant production of HCN (and in some cases
C$_2$H$_2$). It has been suggest that HCN and NH$_3$
will be important disequilibrium constituents on exoplanets with a broad range of
temperatures which should not be ignored in
observational analyses \cite{14MoXXXX.exo}.

\begin{figure}
\begin{center}
\includegraphics[width=.86\textwidth]{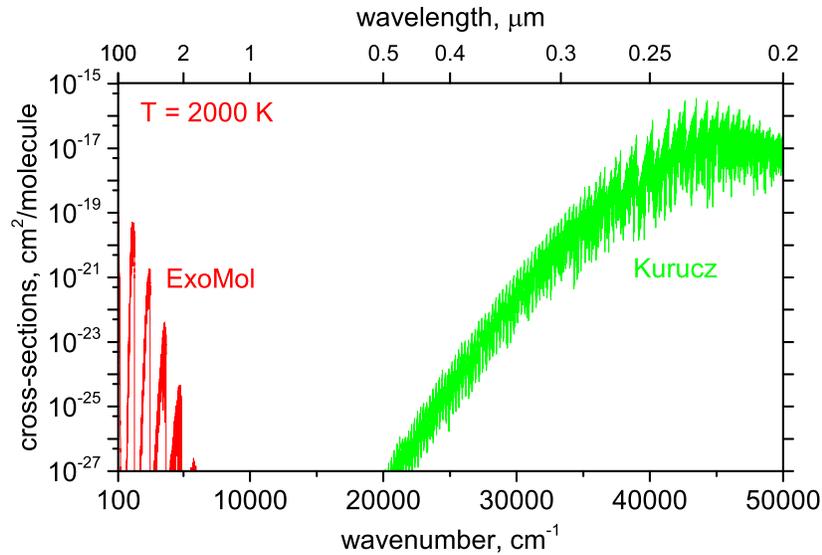}
\caption{SiO absorption at 2000~K: infrared data are taken from ExoMol \protect\cite{jt563} and ultraviolet data from Kurucz \protect\cite{11Kurucz.db}.}
\label{f:SiO}
\end{center}
\end{figure}

Ito \textit{et. al.} \cite{15ItIkMa.exo} suggested that SiO absorption dominates
the UV and IR wavelength regions with the prominent absorption features at around 0.2, 4, 10 and 100 $\mu$m, see Fig.~\ref{f:SiO}.
In particular, in the cases of Kepler-10b and 55 Cnc e, those
features are potentially detectable by the space-based
observations that should be possible in the near future \cite{15ItIkMa.exo}.
  Models suggest that a photon-limited, JWST-class telescope should be able to
detect SiO in the atmosphere of 55 Cnc e with 10 hours of observations
\cite{16KiFeSc} using secondary-eclipse spectroscopy.
Such observations have
the potential to study lava planets even with clouds and lower-atmospheres
\cite{14SaLeRo.exo}.

Other abundant species that may contribute to the transmission spectrum include
CO, OH, and NO at high temperatures. These molecules
should be present in a planet with an O$_2$-rich atmosphere and magma oceans, such
as were recently suggested as the composition of the super-Earth GJ 1132b
\cite{16ScWoZa}.


It is suggested that for atmospheres of hot rocky super-Earths with
high temperature ($>$1800 K) and low pressure almost all rock is vaporized,
while at high pressure ($>$100 bar) much of this material is in the condensed
phase \cite{12ScLoFe.exo}.
Most elements found in rocks are expected to be soluble in steam
\cite{16FeJaWi}, including Mg, Si, and Fe from SiO$_2$-rich (i.e., felsic)
silicates (like Earth's continental crust) and MgO-, FeO-rich (i.e., mafic)
silicates \cite{12ScLoFe.exo}.
This can lead that gases such as Si(OH)$_4$, Mg(OH)$_2$, Fe(OH)$_2$,
Ni(OH)$_2$, Al(OH)$_3$, Ca(OH)$_2$, \mol{NaOH}, and \mol{KOH}
\cite{12ScLoFe.exo}.
Silica (SiO$_2$) dissolves in steam primarily via formation of Si(OH)$_4$
\cite{12Plyasunov}, while  MgO in steam leads to production of gaseous Mg(OH)$_2$,
see, for example, Ref.~\cite{63AlOgLe}. 
However it seems likely that at the temperatures under consideration
many of these more complex species would fragment into diatomic or triatomic species, and
water.

The predicted vaporised constituents of the steam atmosphere at higher
temperatures (4000~K)
include \cite{12ScLoFe.exo}  Fe and \mol{FeO} (products of Fe(OH)$_2$ fragmentation),
\mol{MgO}, Titanium dioxide TiO$_2$ (major Ti-bearing gas with abundance of
1.1\%), PO$_2$ and then \mol{PO} (with
increasing temperature),  MnF$_2$ and MnO (from vaporized bulk Mn), CrO$_2$F,
CrO$_2$, and CrO
(from vaporized bulk CrO), Ca(OH)$_2$ and AlO (although calcium and aluminum are less
abundant).  TiO$_2$ can lead \mol{TiO} \cite{72BaMaGu}, which is well-known to be a source of major absorption
from near-infrared to the optical spectral regions of M dwarfs \cite{aha97}.
There have been several attempts to detect \cite{08DeVide.TiO,08DeVide.TiO} 
and a recent reported detections of TiO in exoplanet atmospheres \cite{jt699}.
Whether complex polyatomic molecules like   Fe(OH)$_2$,
Ca(OH)$_2$,  CrO$_2$F  and  P$_2$O$_5$ will survive at $T > 1000$ K is
questionable.
It should be noted that it is the lower pressure regimes that hold
out the best prospects for analysis using
transit spectroscopy, as the high pressures will tend to result in  opaque
atmospheres.

Post-impact rocky planets are shown to have very similar atmospheric and
therefore spectroscopic properties. According to estimated luminosities, the
hottest post-giant-impact planets will be detectable with near-infrared
coronagraphs on the planned 30 m class telescopes \cite{14LuZaMa}.
The 1-4 $\mu$m region will be most favorable for such observations, offering bright
features and better contrast between the planet and a potential debris disk.
The greenhouse absorbers in a rocky exoplanet atmosphere strongly influence
its cooling properties.  The very large cooling timescales (on the order of
$10^{5}-10^{6}$ yr) lead to the possibility of discovering tens of such
planets in future surveys \cite{14LuZaMa}. It has recently been
suggested \cite{17BaSyxx} that even gas giant planets may form visible
massive, rocky exomoons as a result of giant impacts.

55 Cnc e is currently the most
attractive candidate magma planet for observations
 \cite{jt629,16DeGide}; its atmosphere is amenable
to study using secondary-eclipse
spectroscopy and high-dispersion spectroscopy observations.

It is thought that during its formation of the atmosphere of the early
Earth was dominated by steam which contained water-bearing minerals
\cite{74AlArxx,86MaAbxx,88AbMaxx,88Kasting,88ZaKaPo,08ElSexx.exo,13LeMaCh,16FeJaWi}.
As Lupe {\it et al.} \cite{14LuZaMa} pointed out, modern
state-of-the-art radiative transfer in runaway and near-runaway
greenhouse atmospheres \cite{88AbMaxx,88Kasting} are mainly based on
the absorption of H$_2$O and CO$_2$, with rather crude description of
hot bands and neglecting other opacity sources. It is important,
however, that the line-by-line radiative transfer calculations of
outgoing longwave radiation include greenhouse absorbers of a rocky
exoplanet atmosphere affecting its cooling. Discussion of such data
is given below.

It should be noted that clouds and hazes can lead to flat, featureless
spectra of a super-Earth planet \cite{15MoFoMa.exo}, preventing
detection of some or all of the spectral features discussed above. As
Morley {\it et al} argued \cite{15MoFoMa.exo}, it is however possible
to distinguish between cloudy and hazy planets in emission: NaCl and
sulfide clouds cause brighter albedos with ZnS known to have a
distinct feature at 0.53~$\mu$m.




\begin{table}
\caption{Molecules thought to be
important for spectroscopy of the atmospheres hot rocky super-Earths.}
\begin{tabular}{llll}
\hline
CH$_4$, C$_2$H$_2$,  CO,  CO$_2$,  \\
H$_2$,  HCl, HCN,  HF, H$_2$O, H$_2$S,   \\
KCl, KOH, MgO, Mg(OH)$_2$, \\
NaCl, NaOH, NH$_3$,  NO, OH, PO$_2$,  \\
SiO, SiO$_2$, SO,  SO$_2$,  SO$_3$, ZnS \\
\hline
\end{tabular}\label{species}
\end{table}

A summary of the molecules important for the spectroscopy of hot melting
planets is given in Table~\ref{species}.
The following sections in turn discuss how suitable spectroscopic data
can be assembled and the present availability of such data required for
retrievals from the atmospheres of rocky super-Earths which are
essential for analysis of the exoplanetary observations.
Exactly these types of hot rocky objects will be the likely targets of
NASA's JWST (due for launch in 2018) and other exoplanet transit
observations. Models suggest that magma-planet clouds and
lower-atmospheres can be observed using secondary-eclipse
spectroscopy \cite{16KiFeSc} and that a photon-limited JWST-class
telescope should be able to detect SiO, Na and K in the atmosphere of
55 Cnc e with 10 hours of observations \cite{15ItIkMa.exo}.
Furthermore, albedo
measurements are possible at lower signal to noise; they may correspond to the
albedo of clouds, or the albedo of the surface
\cite{11RoDeDe.exo,14Demory}

\begin{figure}
\begin{center}
\label{f:Snellen}
\includegraphics[width=0.7\textwidth]{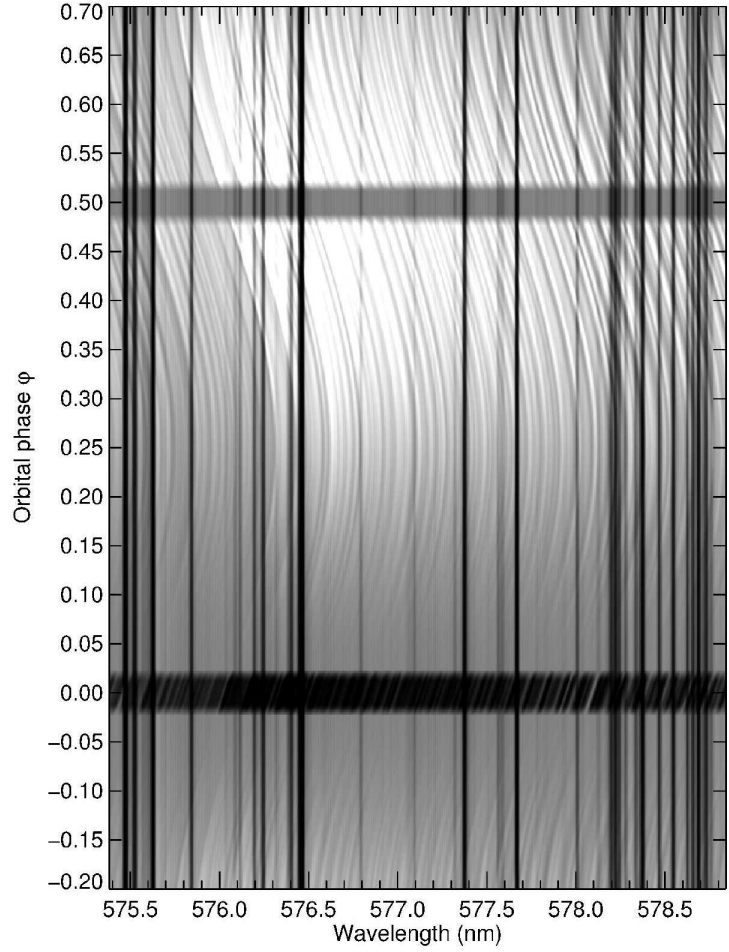}
\caption{Toy model of the phase-dependent Doppler shift of TiO lines along the orbit of HD 209458b. The white curves represent TiO-emission features owing to the inversion layer. The black vertical lines are stellar absorption lines, which are stable in time. This difference in the behaviour of stellar and planetary features provides a means of contrast between star and planet.
Credit: Hoeijmakers {\it et al} \cite{15HoDeSn}, reproduced with permission \copyright\ ESO.}
\end{center}
\end{figure}

High quality is also needed for complementary high-dispersion
spectroscopic \cite{14Snellen,15HoDeSn.TiO,15HoDeSn} (see
Fig.~\ref{f:Snellen}, where the technique is illustrated using Doppler
shifted TiO lines). For example TiO could not be detected in the
optical transmission spectrum of HD 209458b due to (arguably) poor
quality of the TiO spectral data \cite{15HoDeSn}.

The above discussion concentrates on molecular species and infrared spectra.
However, transit observation of atomic spectra at visible
wavelengths, particularly due atomic hydrogen \cite{03VideBa} and
sodium \cite{02ChBrNo}, were actually the earliest spectroscopic
studies of exoplanets. More recently, the Hubble Space Telescope
telescope has been used to perform transit spectroscopy of exoplanets
in the ultraviolet revealing the presence of both neutral
Mg \cite{15BiLaVi} and its ion Mg$^+$ \cite{10FoHaFr}, as well as the possible
detection a variety of
other possible atoms and atomic ions.

\section{Methodology}

The spectroscopic data required to perform atmospheric models and
retrievals comprise line positions, partition functions, intensities,
line profiles and the lower state energies $E"$, which are usually
referenced to as `line lists'.  Given the volume of data required for
construction of such line lists is far from straightforward. When
considering how this is best done it is worth dividing the systems
into three classes:
\begin{enumerate}
\item Diatomic molecules which do not contain a transition metal
atom which we will class as simple diatomics;
\item Transition metal containing diatomics such as TiO;
\item Polyatomic molecules.
\end{enumerate}

For simple diatomics it is possible to construct experimental line lists
which cover the appropriate ranges in both lower state energies and wavelength. There are  line lists available which are
based entirely on direct use of experimental data \cite{14YuDrMi.O2} or
use of empirical energy levels and calculated, \ai, dipole
moments and hence transition intensities \cite{16BrBeWe.OH}. It is
also possible to generate such line lists by direct solution of
the nuclear motion Schr\"{o}dinger equation \cite{level,jt609}
for a given potential
energy curve and dipole moment function \cite{jt529}. This means that
while there are still simple diatomics for which line lists are needed,
it should be possible to generate them in a reasonably
straightforward fashion.

\begin{figure}
\begin{center}
\includegraphics[width=0.95\textwidth]{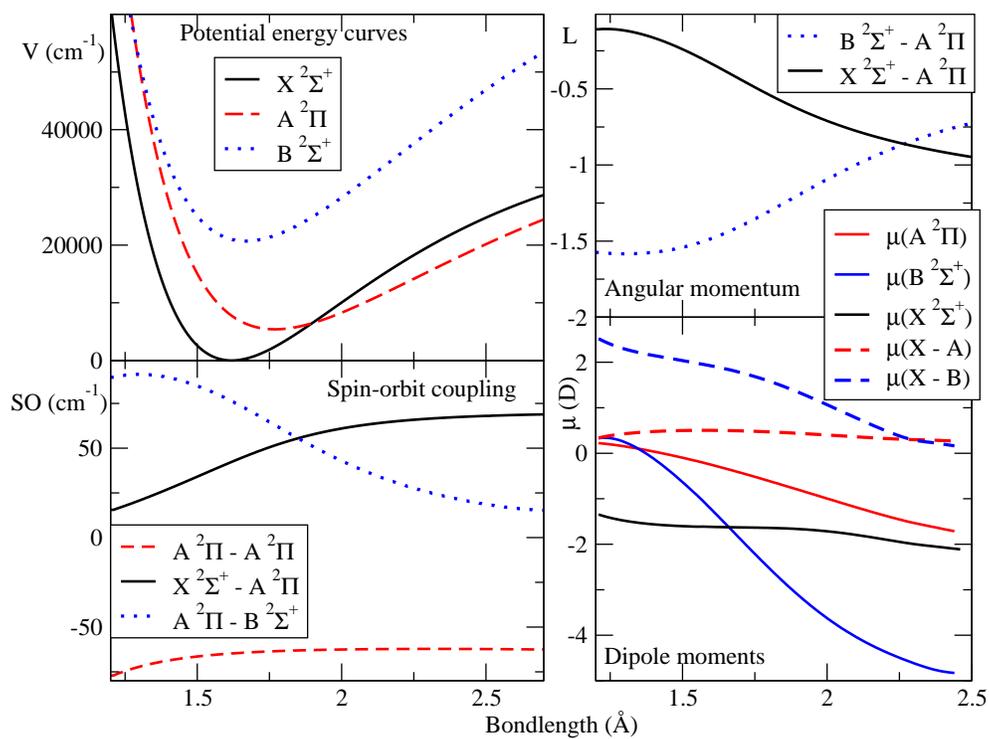}
\caption{Curves representing the spectroscopic model of AlO \cite{jt589,jt598}; in this
model the potential energy curves are coupled by both spin-orbit and electronic angular momentum effects.}
\label{f:AlO:curves}
\end{center}
\end{figure}

When the diatomic contains a transition metal, things are much less
straightforward \cite{jt632,jt623}.  These systems have low-lying electronic states and
it is necessary to consider vibronic transitions between several
states plus couplings and transition dipole moments
between the states. The curves required to give
a full spectroscopic model of systems for which vibronic transitions
are important are summarized in Fig~\ref{f:AlO:curves} for aluminium
monoxide, AlO. AlO is a relatively simple system which only requires
consideration of three electronic states. This should be contrasted
with the yet unsolved case of iron monoxide, FeO, where there are more
than fifty low-lying electronic states \cite{11SaMiMa.FeO} which means
that a full spectroscopic model will require consideration of
several hundred coupling curves and a similar number of transition
dipoles.

Experimentally, open shell transition metal systems are challenging to prepare and
the resulting samples are usually not thermal which makes it hard to obtain
absolute line intensities. Under these circumstances it is still possible
to measure decay lifetimes which are very useful for validating theoretical models.
Lifetime meaurements are currently rather rare and we  would encourage
experimentalists to make more of these for transition methal systems.
Furthermore, the many low-lying electronic
states are often strongly coupled and interact, which makes it difficult to
construct robust models of the experimental data. From a theoretical
perspective, the construction of reliable potential
energy curves and dipole moment functions remains difficult with currently
available \ai\ electronic structure
methods \cite{jt632,jt623}. The result is that even for
important systems such as TiO \cite{00AlHaS1.TiO}, well-used line lists
\cite{98Plxxxx.TiO,98Scxxxx.TiO} are known to be inadequate \cite{15HoDeSn.TiO}.

For polyatomic molecules there have been some attempts to construct
line lists directly from experiment, for example for ammonia \cite{11HaLiBe,12HaLiBe.NH3} and methane \cite{12HaBeMi.CH4,15HaBeBa.CH4}. However,
this process is difficult and can suffer from problems with both
completeness \cite{jt542} and the correct inclusion of
temperature dependence.
The main means of constructing  line lists for these systems has therefore
been variational nuclear motion calculations.

There are three groups who are systematically producing extensive
theoretical line lists of key astronomical molecules.  These are the
NASA Ames group of Huang, Lee and Schwenke
\cite{14HuGaFr.CO2,16HuScLe.SO2}, the Reims group of Tyuterev, Nikitin
and Rey who are running the TheoReTS project \cite{TheoReTS} and our
own ExoMol project \cite{jt528,jt631}. While there are differences in
detail, the methodologies used by these three groups are broadly
similar. Intercomparison for molecules such as SO$_2$, CO$_2$ and
CH$_4$, discussed below, are generally characterized by good overall
agreement between the line lists presented by different groups with
completeness and coverage being the main features to distinguish them.
Thus, for example, both the TheoReTS and ExoMol groups pointed out
that the 2012 edition of the HITRAN database \cite{jt557} contained
a spurious feature due to methane near 11 $\mu$m \cite{14ReNiTy.CH4,jt564},
which led to its removal in the 2016 release of HITRAN \cite{jt691}.

\begin{figure}
\begin{center}
\includegraphics[width=0.9\textwidth]{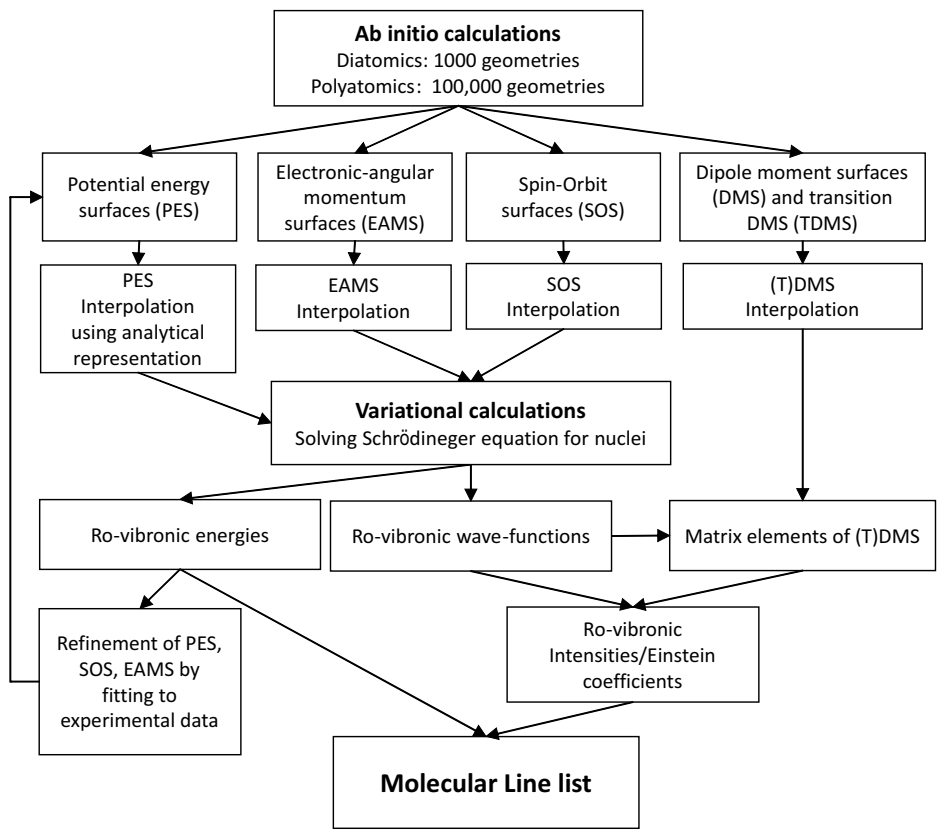}
\caption{ The computational recipe used to produce a line list of rotation-vibration-electric (rovibronic) transitions using a mixture if \ai\ and  variational
nuclear motion calculations.
\label{f:procedure} }
\end{center}
\end{figure}

Figure~\ref{f:procedure}  illustrates the procedure whereby
line lists of both rotation-vibration and rotation-vibration-electronic transitions are computed using variational
nuclear motion calculations.
These calculations are based on the direct use of a potential energy
surface (PES) to give energy levels and associated wavefunctions, and
dipole moment surfaces (DMS) to give transition intensities
\cite{jt475}. For vibronic spectra such as those encountered with the open-shell diatomics the spin-orbit (SO), electronic
angular momentum (EAM) and transition dipole moments (TDM) curves are also required.
The procedure is well established \cite{jt511} in that
for all but a small number of systems with very few electrons \cite{jt347,jt478,jt666},
the PES used is spectroscopically determined. That is, an initial high-accuracy
{\it ab initio} PES is systematically adjusted until it reproduces
observed spectra as accurately as possible. Conversely, all the evidence
suggests that the use of a purely \ai\ DMS gives better results than
attempts to fit this empirically \cite{jt156,jt509,jt613}.

The PE, SO, EAM and (T)DM surfaces are usually interpolated by
appropriate analytical representations to be used as an input for the
nuclear motion program. The quality of the PES (as well as of the
coupling curves) is improved \textit{a priori} by refining the
corresponding expansion parameters by comparison with laboratory high
resolution spectroscopic data. This refinement, particularly of PESs,
using spectroscopic data is now a well-developed procedure pursued by
many groups. For example, the Ames group have provided a number highly
accurate PES for small molecules based on very extensive refinement of
the PES \cite{11HuScLe.NH3,12HuScTa.CO2,14HuScLe.SO2} starting from
initial, high accuracy, {\it ab initio} electronic stucture
calculations. Our own preference is to constrain such fits to remain
close to the original {\it ab initio} PES \cite{jt503}; this has the benefit of forcing
the surface to remain physically correct in regions not well-characterized
experimentally. Such regions are often important for calculations
of extensive, hot line lists. Further discussion of the methods
used to refine PESs can be found in Ref. \cite{jt511}.

Our computational tools include the variational nuclear-motion programs Duo
\cite{jt609}, DVR3D \cite{jt338}, and TROVE \cite{TROVE,15YaYuxx.method}
which calculate
the rovibrational energies, eigenfunctions, and transition dipoles
for diatomic, triatomic and larger polyatomic molecules, respectively.
These programs have proved capable of producing accurate spectra
for high rotational excitations and thus for high-temperature
applications.  All these codes have been adapted to face
the heavy demands of computing very large line lists \cite{jt626}
and are available as freeware.

Duo was recently developed especially  for treating
open-shell system of astrophysical importance
\cite{jt598,jt599,jt618,jt644}.  To our knowledge Duo is currently the
only code capable of generating spectra for general diatomic molecules
of arbitrary number and complexity of couplings.

DVR3D~\cite{jt338}
was used to produce line lists for several key triatomics, including
H$_2$S, SO$_2$, H$_2$O, CO$_2$, HCN
\cite{jt640,jt635,jt625,jt667,jt678,jt665,jtpoz,jt298}.
DVR3D is capable of treating ro-vibrational states up to
dissociation and above \cite{jt443}. A new version appropriate
for the calculation of fully-rotationally resolved electronic
spectra of triatomic species has just been developed and tested
for the X -- A band in SO$_2$ \cite{jt697}.

TROVE is a general polyatomic code that has been used to generate
line lists for hot NH\3, PH\3, H\2CO, HOOH, SO\3, CH\4\
\cite{jt500,jt571,jt592,jt597,jt620,jt641,jt698}. Intensities in TROVE are
computed using the new code GAIN \cite{jt653} which was written and
adapted for graphical
processing units (GPUs) to compute Einstein coefficients (or oscillator
strengths) and integrated absorption coefficients for all individual
rotation-vibration transitions at different temperatures.  Given the
huge number of transitions anticipated to be important at elevated
temperatures, the usage of GPUs provides a huge advantage.
However TROVE requires special adaptation \cite{jtlinTROVE} to treat
linear molecules such as the astronomically important acetylene (HCCH).

An alternative theoretical procedure
has been used by Tashkun and Perevalov from Tomsk. Their methodology
uses effective Hamiltonian fits to experimental data for both energy
levels and transition dipoles. This group has provided high-temperature
line lists for the linear CO$_2$ \cite{11TaPe.CO2} molecule and
the NO$_2$ \cite{16LuLaDu.NO2} system. This
methodology reproduces the positions of observed lines to much higher
accuracy than the variational procedure but generally extrapolates
less well for transitions involving states which are outside the range
of those that have been observed in the laboratory. In particular,
comparisons with high-resolution transmission measurements of CO$_2$
at high temperatures for industrial applications suggest that
indeed the CDSD-4000 CO$_2$ line list loses accuracy at higher temperatures.
We note that the Ames group have produced variational line lists
for CO$_2$ designed to be valid up to 1500 K \cite{13HuFrTa.CO2}
and 4000~K \cite{17HuScFr.CO2}.

As mentioned above, a disadvantage of the use of variational nuclear
motion calculations is that the transition frequencies are rarely
predicted with spectroscopic accuracy. One method of rectifying this
problem is by use of the MARVEL (measured active
rotational-vibrational energy levels) procedure
\cite{jt412,12FuCs.method}. The MARVEL procedure inverts the measured
transition frequencies to provide energy levels from which not only
can the original transition frequencies be regenerated but all other
transitions linking these states can also be obtained with
experimental accuracy. However, the MARVEL procedure does not provide any
information on levels which have yet to be observed experimentally.
MARVEL datasets of energy levels are available for a range of
astronomically important molecules including water
\cite{jt562,jtwaterupdate}, H$_3^+$ \cite{13FuSzFa.H3+,13FuSzMa.H3+},
NH$_3$ \cite{jt608,jtNH3update}, C$_2$ \cite{jt637}, TiO \cite{jt672} 
and HCCH \cite{jtC2H2Marvel}.  In particular, the
energy levels and transition frequencies from the analysis of TiO
spectra should provide the high-resolution transition frequencies need
to allow the detection of TiO in exoplanets using high-dispersion
spectroscopy for which previously available laboratory data was not
precise enough \cite{15HoDeSn.TiO}. Indeed this analysis pointed
to a number of issues with previous analysis of observed TiO spectra
and significant shifts in transition frequencies compared to those
provided by the currently
available line lists \cite{98Plxxxx.TiO,98Scxxxx.TiO}.

The MARVEL energy levels can also be used to replace computed ones
in line lists. This has already been done for several line lists
\cite{jt615,jt666,jt686}. This process is facilitated by the
ExoMol data structure \cite{jt548,jt631} which does not store transition frequencies
but instead computes them  from a states file containing all the energy levels.
This allows changes of the energy levels at the end of the calculation
or even some time later \cite{jt374,jt570} should improved energy levels
become available.

The polyatomic molecules discussed above are all closed shell species.
However the open shell species PO$_2$, mentioned above, and
CaOH  are thought to be important for hot atmospheres \cite{09Bexxxx.exo}.
There have been a number of variational nuclear motion
calculations on the spectra
of open shell triatomic systems \cite{02JeOkKr,07BuKrYu,07OdMeJe,08HeOkNa,13MaMeHe,16OsScBu},
largely based on the use of Jensen's MORBID approach \cite{95JeBrKr.method}. However,
we are unaware of any extensive line lists being produced for such systems.
The extended version of DVR3D \cite{jt697} mentioned above should, in due course, be
applicable to these problems.

For closed-shell polyatomic molecules, such as NaOH, KOH, SiO$_2$,
for which spectra involve rotation-vibration transitions on the ground
electronic state,
one would use a standard level of \ai\ theory such as
CCSD(T)-f12/aug-cc-pVTZ on a large grid of geometries ($\simeq
10,000$) to compute both the PES and DMS.  For diatomic molecules (NaH,
KCl, SiO, MgO, ZnS, SO) characterized by multiple interacted curves the multi-reference
configuration interaction (MRCI)
method in conjunction with the aug-cc-pVQZ or higher basis sets
is a reasonable choice, with relativistic and core-correlation effects
included where feasible.  The potential energy and coupling curves should then be
optimized by fitting to the experimental energies or transitional
wavenumbers. Indeed where there is a large amount of experimental
data available, then the choice of initial potential energy curves becomes
almost unimportant \cite{jt563}. However, the \ai\ calculation of good dipole
curves is always essential since these are not in general tuned to observation.

The ExoMol line lists are prepared so that they can easily be
incorporated in radiative transfer codes \cite{jt631}.  For example,
these data ar directly incorporated into the UCL Tau-REx retrieval code
\cite{jt545,jt593,jt611}, a radiative transfer model for transmission,
emission and reflection spectroscopy from the ultra-violet to infrared
wavelengths, able to simulate gaseous and terrestrial exoplanets at
any temperature and composition.  Tau-REx uses the linelists from
ExoMol, as well as HITEMP \cite{jt480} and HITRAN \cite{jt557} with
clouds of different particle sizes and distribution, to model
transmission, emission and reflection of the radiation from a parent
star through the atmosphere of an orbiting planet. This allows
estimates of abundances of absorbing molecules in the atmosphere, by
running the code for a variety of hypothesised compositions and
comparing to any available observations. Tau-REx is mostly based on
the opacities produced by ExoMol with the ultimate goal to build a
library of sophisticated atmospheres of exoplanets which will be made
available to the open community together with the codes.  These models
will enable the interpretation of exoplanet spectra obtained with
future new facilities from space \cite{jt606,jt627} and the ground (VLT-SPHERE, E-ELT,
JWST).

Of course there are a number of other models for exoplanets and
similar objects which rely on spectroscopic data as part of their
inputs.  These include modelling codes such as NEMESIS
\cite{08IrTeKo.model}, BART \cite{16BlHaCu}, CHIMERA \cite{CHIMERA} and a recent adaption of the UK Met Office
global circulation model (GCM) called ENDGame
\cite{14MaBaAc,16AmMaBa}. More general models such as VSTAR
\cite{12BaKexx.dwarfs} are designed to be applied to spectra of
planets, brown dwarfs and cool stars. The well-used BT-Settl
brown-dwarf model \cite{12AlHoFr,BT-Settl} can also be used for
exoplanets.  There are variety of other brown dwarfs
\cite{06BuSuHu.dwarfs} and cool star models
\cite{08Tsuji,08GuEdEr.model,14Kurucz}. These are largely concerned
with the atmospheres of the hydrogen rich atmospheres which are, of
course, characteristic of hot Jupiter and hot Neptune exoplanets,
brown dwarfs and stars.

Besides direct input to models, line lists are used
to provide opacity functions \cite{07ShBuxx.dwarfs,08FrMaLo.exo,11Kurucz.db,14Bernath,14FrLuFo}
whose reliability are well-known to be limited by the availability of
good underlying spectroscopic data \cite{08CuMaSa.dwarfs}.
Cooling functions for key molecules are also important for the
description of atmospheric processes in hot rocky objects. These functions
are straightforward to compute from a comprehensive line lists \cite{jt624};
this  involve computation of integrated emissivities from all lines on a
grid of temperatures typically ranging between 0 to 5000~K.

\section{Available spectroscopic data}

Spectroscopic studies of the Earth's atmosphere are supported by
extensive and constantly updated databases largely comprising
experimental laboratory data \cite{jt557,jt636}. Thus for earth-like
planets, by which we mean rocky exoplanets with an atmospheric
temperature below 350~K, the HITRAN database \cite{jt691} makes a good
starting point.  However, at higher temperatures datasets designed for
room temperature studies rapidly become seriously incomplete
\cite{jt572}, leading to both very significant loss of opacity and
incorrect band shapes. The strong temperature dependence of the
various molecular absorption spectra is illustrated in figures given throughout this review
which compare simulated absorption spectra at 300 and 2000 K for
key species.

HITRAN's sister database, HITEMP, was developed
to address the problem of high temperature spectra.  However the latest release of HITEMP
\cite{jt480} only contains data on five molecules, namely CO, NO,
O$_2$, CO$_2$ and H$_2$O. For all these species there are more recent
hot line lists available which improve on the ones presented in
HITEMP. These line lists are summarised in Table~\ref{tab:exodata}
below.

Table~\ref{species} gives a summary of species suggested by the
chemistry models as being important in the atmospheres of hot super-Earths.
Spectroscopic line lists are already available for many of
the key species.   Most of the species suggested by the
chemistry models of such objects are already in the ExoMol database, which includes
line list taken from sources other than the ExoMol project itself. This
includes H$_2$O, CH$_4$, NH$_3$, CO$_2$, SO$_2$ (see a synthetic spectrum of
water at high temperature and pressure).
Line lists for
other important species, such as NaOH, KOH, SiO$_2$,  PO, ZnS and SO  are currently missing.
Table~\ref{tab:exodata} presents a summary of line lists available for
atmospheric studies of hot super-Earths.

Line lists for some diatomics are only partial: for example accurate
infrared (rotation-vibration) line lists exists for CO, SiO, KCl,
NaCl, NO, but none of these line lists consider shorter-wavelength,
vibronic transitions which lie in the near-infrared (NIR), visible (Vis) or
ultraviolet (UV), depending on the species concerned.
 NIR will be covered by the NIRSpec
instrument on the board of JWST only at lower resolution and therefore the
completeness of the opacities down to 0.6~$\mu$m will be crucial for the
atmospheric retrievals.  Such data, when available,  will be important for the
interpretation of present and future exoplanet spectroscopic observations.

\begin{table}
\caption{spectroscopic line lists available for studies of the atmospheres hot
super-Earth exoplanets}
\label{tab:exodata}
\begin{center}
\tabcolsep=3pt
\footnotesize
\resizebox{0.95\textwidth}{!}{%
\begin{tabular}{l@{\footnotesize}crcrlll}
\hline\hline
Molecule&$N_{\rm iso}$&$T_{\rm max}$&$N_{elec}$&$N_{\rm
lines}$&DSName&Reference&Methodology\\
\hline
SiO&5&9000&1& 254~675&EJBT&Barton \ea\ \cite{jt563}&ExoMol\\
MgH&1&&3& 30~896&&GharibNezhad \ea\ \cite{13GhShBe.MgH}&Empirical\\
CaH&1& &2&6000 && Li {\it et al} \cite{11LiHaRa.CaH}&Empirical\\
NH&1&&1&10~414&&Brooke \ea\ \cite{14BrBeWe.NH}&Empirical\\
CH&2&&4&54~086&&Masseron \ea\ \cite{14MaPlVa.CH}&Empirical\\
CO&9&9000&1&752~976&&Li {\it et al} \cite{15LiGoRo.CO}&Empirical\\
OH&?&6000&1&$\Delta v = 13$&&Brooke \ea\ \cite{16BrBeWe.OH}&Empirical\\
CN&1&&1&195~120&&Brooke \ea\ \cite{14BrRaWe.CN}&Empirical\\
CP&1&&1&28~735&&Ram \ea\ \cite{14RaBrWe.CP}&Empirical\\
HF&2&&1&13~459&&Li \ea\ \cite{13LiGoHa.HCl}&Empirical\\
HCl&4&&1&34~250&&Li \ea\ \cite{13LiGoHa.HCl}&Empirical\\
NaCl&2&3000&1& 702~271 &Barton&Barton \ea\ \cite{jt583}&ExoMol\\
KCl&4&3000&1& 1~326~765  &Barton&Barton \ea\ \cite{jt583}&ExoMol\\
PN&2&5000&1&142~512&YYLT&Yorke \ea\ \cite{jt590}&ExoMol\\
AlO&4&8000&3&4~945~580&ATP& Patrascu \ea\ \cite{jt598}&ExoMol\\
NaH&2&7000&2&79~898&Rivlin&Rivlin \ea\ \cite{jt605}&ExoMol\\
CS&8&3000&1&548~312&JnK&Paulose \ea\ \cite{jt615}&ExoMol\\
CaO&1&5000&5&21~279~299&VBATHY&Yurchenko \ea\ \cite{jt618}&ExoMol\\
NO& 6&5000&5&2~281~042&NOname&Yurchenko \ea\ \cite{jt686}&ExoMol\\
VO&1&5000&13&277~131~624&VOMYT& McKemmish \ea\ \cite{jt644}&ExoMol\\
H$_2$O &4$^a$&3000&1&12~000~000~000&PoKaZoTeL& Polyansky \ea\ \cite{jtpoz}&ExoMol\\
CO$_2$&4$^b$& 4000&1&628,324,454& CDSD--4000&Tashkun \&\ Perevalov \cite{11TaPe.CO2}&Empirical\\
SO$_2$&1&2000&1&1~300~000~000&ExoAmes& Underwood \ea\ \cite{jt635}&ExoMol\\
H$_2$S&1&2000&1&115~530~373&ATY2& Azzam \ea\ \cite{jt640}&ExoMol\\
HCN/HNC&2$^c$&4000&1&399~000~000&Harris& Barber \ea\ \cite{jt570}&ExoMol\\
NH$_3$&2$^d$&1500&1&1~138~323~351&BYTe& Yurchenko \ea\ \cite{jt500}&ExoMol\\
PH$_3$&1&1500&1&16~803~703~395&SAlTY& Sousa-Silva \ea\ \cite{jt592}&ExoMol\\
CH$_4$&1&1500&1&9~819~605~160&10to10& Yurchenko \&\ Tennyson \cite{jt564}&ExoMol\\
\hline
\hline
\end{tabular}
}
\linebreak
{\footnotesize
\hfill
\noindent
$N_{\rm iso}$: Number of isotopologues considered; $T_{\rm max}$: Maximum temperature for which the line list is complete; $N_{elec}$: Number of electronic states considered; $N_{\rm lines}$:  Number of lines: value is for the main isotope. DSName: Name of line list chosen by the authors, if applicable.
$^a$ The VTT line list for HDO due to Voronin \ea\ \cite{jt469} and HotWat78
due to Polyansky \ea\ \cite{jt665} for H$_2$$^{17}$O and H$_2$$^{18}$O are also
available. $^b$ Very recently Huang {\it et al} \cite{17HuScFr.CO2}
 have computed the Ames-2016 line lists
for 13 isotopologues of CO$_2$ which also extend to 4000 K.
$^c$ A line list for H$^{13}$CN/HN$^{13}$C due to Harris \ea\ \cite{jt447} is
also available.
$^d$ There is a room temperature $^{15}$NH$_3$ line list due to Yurchenko
\cite{15Yurche.NH3}.
}

\end{center}

\end{table}

Below we consider the status of spectroscopic data for key molecules
in turn.

\begin{figure}
\begin{center}
\includegraphics[width=.86\textwidth]{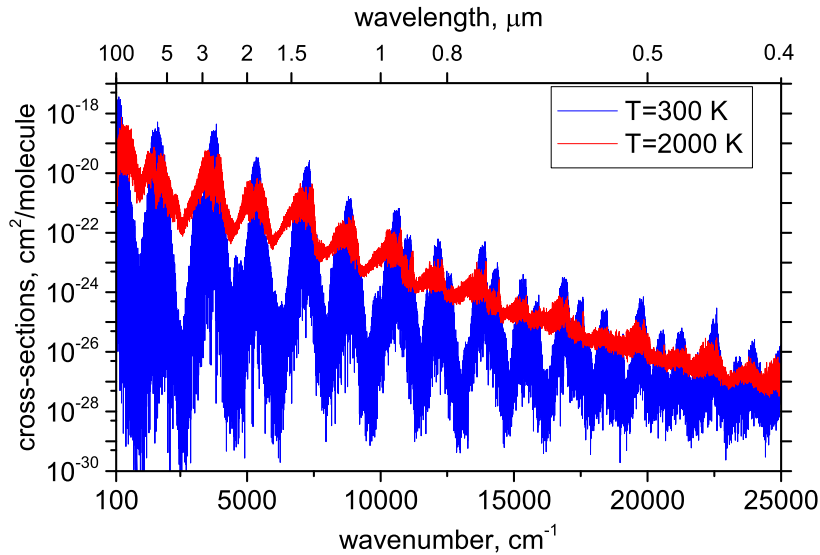}
\caption{Absorption spectrum of H$_2$O at $T$ =300~K and 2000~K
simulated using the  POKAZaTEL line list \cite{jtpoz}.}
\label{f:H2O}
\end{center}
\end{figure}

{\bf H$_2$O}: As discussed above, water is the key molecule in the
atmospheres of rocky super-Earths.  There are a number of published
water line lists available for modelling hot objects
\citep{jt143,92WaRoxx.CO2,jt197,97PaScxx.H2O,JJS01,sp00,jt378,jt480}.
Of these the most widely used are the Ames line list of Partridge and
Schwenke \cite{97PaScxx.H2O}, or variants based on it, and the BT2
line list \cite{jt378}, which provided the basis for water in the
HITEMP database \cite{jt480} and the widely-used BT-Settl brown dwarf
model \cite{07AlAlHo.model}.  The Ames line list is more accurate than
BT2 at infra red wavelengths but less complete meaning that it is less
good at modelling hotter objects. Recently Polyansky \ea\ have
computed the POKAZaTEL line list \cite{jtpoz} which is both more
accurate and more complete than either of these. We recommend the use
of this line list, which is illustrated in Fig.~\ref{f:H2O}, in future studies.

\begin{figure}
\begin{center}
\includegraphics[width=.86\textwidth]{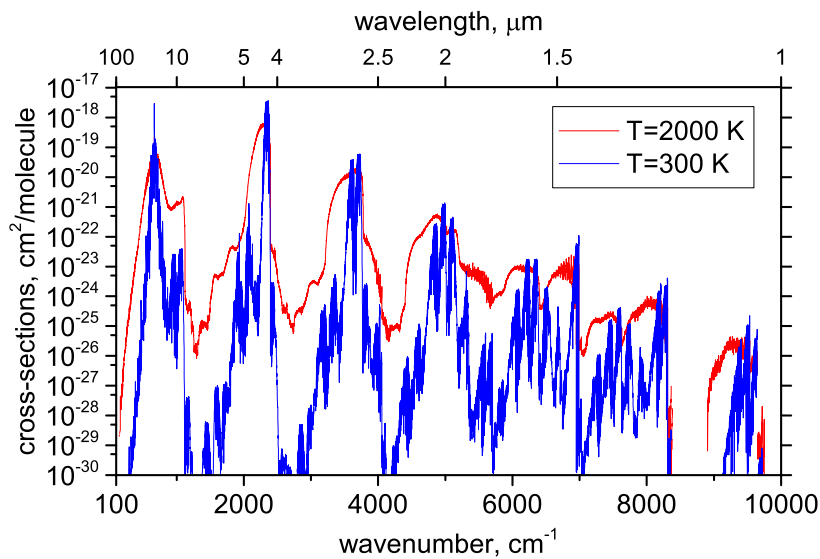}
\caption{Absorption spectrum of CO$_2$ at $T$ =300~K and 2000~K
simulated using HITEMP \cite{jt480}.}
\label{f:CO2}
\end{center}
\end{figure}

{\bf CO$_2$}: Again there are number of line lists available for hot
CO$_2$. In particular Taskun and Perevalov distribute these via their
carbon dioxide spectroscopic databank (CDSD)
\cite{02TaPeTe.CO2,11TaPe.CO2}, an early version of CDSD formed the
input for HITEMP. The Ames group produced a variational line list
valid up to 1500 K \cite{13HuFrTa.CO2}. Recent work on CO$_2$ has
improved computed transition intensities to point where they as
accurate as the measured ones \cite{jt613,jt625}; this suggests that
there is scope for further improvement in hot line lists for this
system; some work in this direction has recently been undertaken by
Huang {\it al al} \cite{17HuScFr.CO2}.  Fig.~\ref{f:CO2} illustrates
the temperature-dependence of the CO$_2$ absorption spectrum in the
infrared.

\begin{figure}
\begin{center}
\includegraphics[width=.86\textwidth]{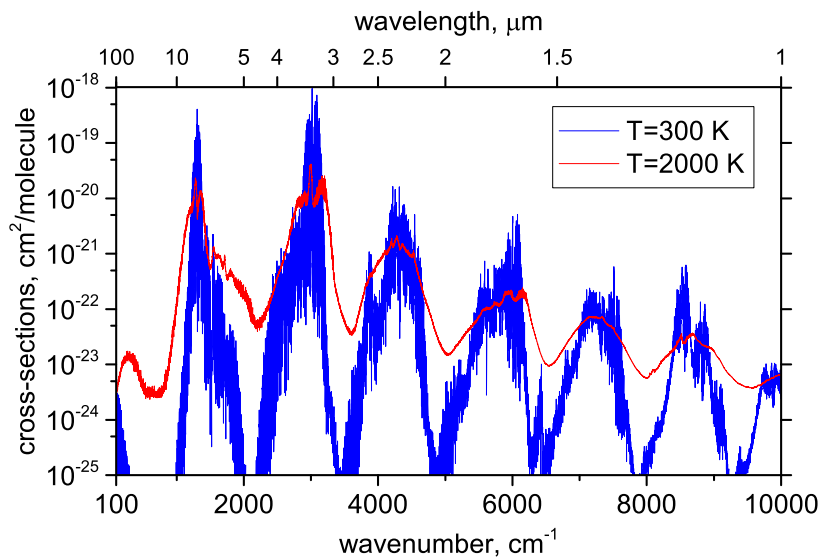}
\caption{Absorption spectrum of CH$_4$ at $T$ =300~K and 2000~K
simulated using the 10to10 line list \cite{jt564}.}
\label{f:CH4}
\end{center}
\end{figure}

{\bf  CH$_4$}: methane is an important system in carbon-rich atmospheres and the
construction
of hot methane line lists has been the subject of intense recent study by a
number of groups both
theoretically
\cite{01ScPaxx.CH4,02Scxxxx.CH4,09WaScSh.CH4,13BaWeSu.CH4,13WaCaxx.CH4,
13MiChTr.CH4,13ReNiTy.CH4,13ReNiTy.CH4.i,jt572,jt564,14ReNiTy.CH4}
and experimentally \cite{12HaBeMi.CH4,15HaBeBa.CH4}. The most complete line
lists currently available are our 10to10 line
list \cite{jt564}, which is very extensive but only valid below 1500 K, and the
Reims line list \cite{14ReNiTy.CH4}, which spans a reduced wavelength
range but is complete up to 2000 K. In fact we  extended 10to10 to higher
temperature some time ago but the result
is a list of 34 billion lines which is unwieldy to use. We have therefore been
working data compaction techniques
based on the use of either background, pressure-independent cross sections
\cite{15HaBeBa.CH4} or super-lines \cite{TheoReTS}. This line list will be released shortly \cite{jt698}. Figure~\ref{f:CH4} illustrates
the temperature-dependence of the methane absorption spectrum in the infrared. The strongest bands are at 3.7 and 7.7~$\mu$m.

{\bf SO$_{2}$ and SO$_3$}: A  number of line list for SO$_2$ have been computed
by the  Ames group \cite{14HuScLe.SO2,15HuScLe.SO2,16HuScLe.SO2};
the most compressive is one produced in collaboration between ExoMol and Ames
\cite{jt635}, see Fig.~\ref{f:SO2}. This line list was validated using experimental data recorded
at the Technical University of Denmark (DTU). ExoMol have also provided
line lists for SO$_3$ \cite{jt554,jt641}. The largest of these, appropriate
for temperatures up to 800 K, contains 21 billion lines. However, validation
of this line list against experiments performed at DTU points to significant differences
in the line intensities, suggesting that more work is required on the SO$_3$
dipole moment.

\begin{figure}
\begin{center}
\includegraphics[width=.86\textwidth]{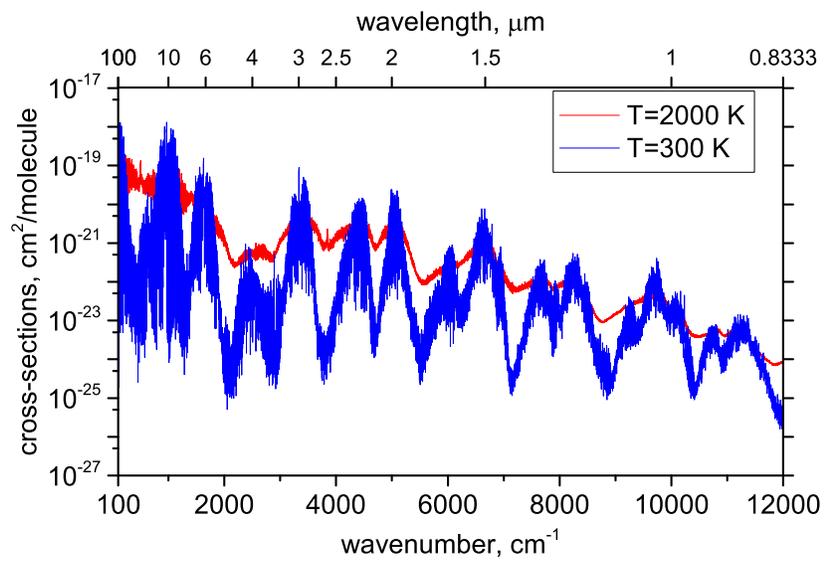}
\caption{Absorption spectrum of NH$_3$ at $T$ =300~K and 2000~K simulated
using the line list BYTe \cite{jt500}.}
\label{f:NH3}
\end{center}
\end{figure}

{\bf NH$_{3}$}: Ammonia has a very prominent absorption feature at about 10~$\mu$m.
Extensive line lists for ammonia are available
\cite{jt466,jt500}. The BYTe
line list \cite{jt500}, which was explicitly designed for needs of exoplanet
spectroscopy in
mind, has been used to model
spectra of brown dwarfs \cite{12BaKexx.dwarfs,jt484,jt596}. However, BYTe loses
accuracy
in the near infrared. Rather old laboratory measurements of room temperature
 for ammonia have recently
been assigned \cite{jt633,jt683}. These data plus improved \ai\ treatment of the
problem \cite{jt634} and a MARVEL analysis leading to a set of accurate, empirical
energy levels \cite{jt608,jtNH3update}
will form the basis of a new line list which will both extend the range and
improve on the accuracy of BYTe. Fig.~\ref{f:NH3} illustrates the absorption spectra of ammonia
at $T=$ 300~K and 2000~K. The strongest and most prominent feature is at 10~$\mu$m.

\begin{figure}
\begin{center}
\includegraphics[width=.86\textwidth]{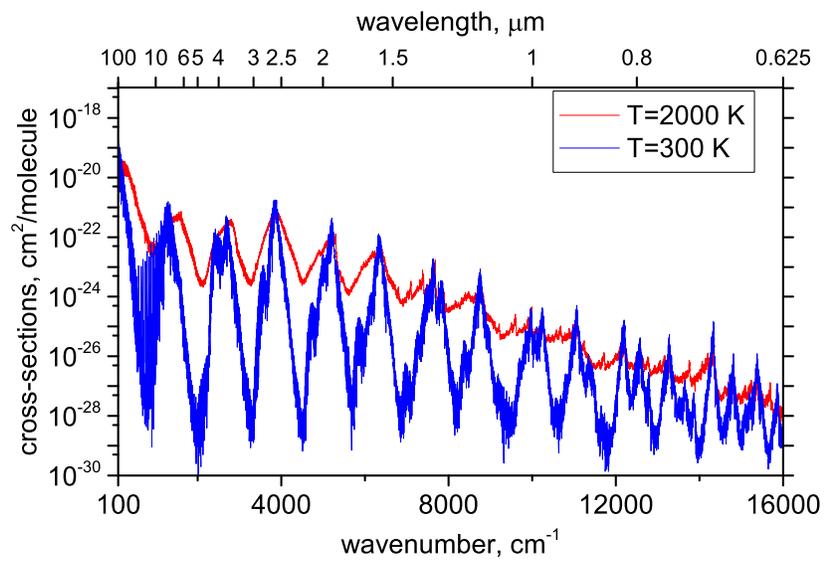}
\caption{Absorption spectrum of H$_2$S at $T$ =300~K and 2000~K simulated
using the ExoMol line list AYT2 \cite{jt640}.}
\label{f:H2S}
\end{center}
\end{figure}

{\bf H$_{2}$S}: The main source of the emission of H$_2$S on Earth is from life \cite{00Watts}.
It has been, however, ruled out as a potential biosignature in atmospheres of exoplanets \cite{13HuSeBa}. H$_2$S is also generated by volcanism.
Fig.~\ref{f:H2S} illustrates the absorption spectra of H$_2$S
at $T=$ 300~K and 2000~K based on the AYT2 line list  \cite{jt640}.

\begin{figure}
\begin{center}
\includegraphics[width=.86\textwidth]{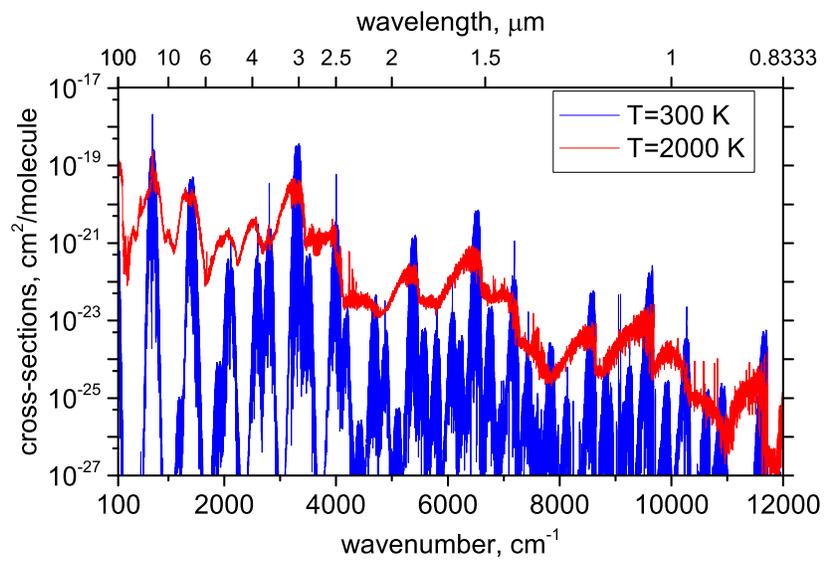}
\caption{Absorption spectrum of the HCN/HNC system
at $T$ =300~K and 2000~K simulated using the ExoMol line list \cite{jt570}.}
\label{f:HCN}
\end{center}
\end{figure}

{\bf HCN}: Line lists for hydrogen cyanide were some of the first
calculated using variational nuclear motion calculations
\cite{84ErGuJo.HCN,jt298}. Indeed the first of these line list was the
basis of a ground-breaking study by J{\o}rgensen
\etal~\cite{85JaAlGu.HCN} showed that use of a comprehensive HCN line
list in a model atmosphere of a `cool' carbon star made a huge
difference: extending the model of the atmosphere by a factor of 5,
and lowering the gas pressure in the surface layers by one or two
orders of magnitude. The line list created and used by J{\o}rgensen
and co-workers \cite{84ErGuJo.HCN,85JaAlGu.HCN} only considered HCN.
However HCN is a classic isomerizing system and the HNC isomer should
be thermally populated at temperatures above about 2000~K
\cite{jt304,jt321}. More recent line lists
\cite{jt298,jt374,jt447,jt570} consider both HCN and HNC together.  All these line lists are based on the
use of \ai\ rather than spectroscopically-determined PESs, which can
lead to significant errors in the predicted transition frequencies
\cite{jt283}. However the most recent line list, due to Barber \ea\
\cite{jt570} used very extensive sets of experimental energy levels
obtained by Mellau for both hot HCN and hot HNC
\cite{11Mexxxx.HCN,11Mexxxx.HNC} to improve predicted frequencies to,
essentially, experimental accuracy. This line list was used for the
recent, tentative detection of HCN on super-Earth 55 Cancri e
\cite{jt629}. The line list of Barber \ea\
\cite{jt570} is illustrated in Fig.~\ref{f:HCN}.

\begin{figure}
\begin{center}
\includegraphics[width=.86\textwidth]{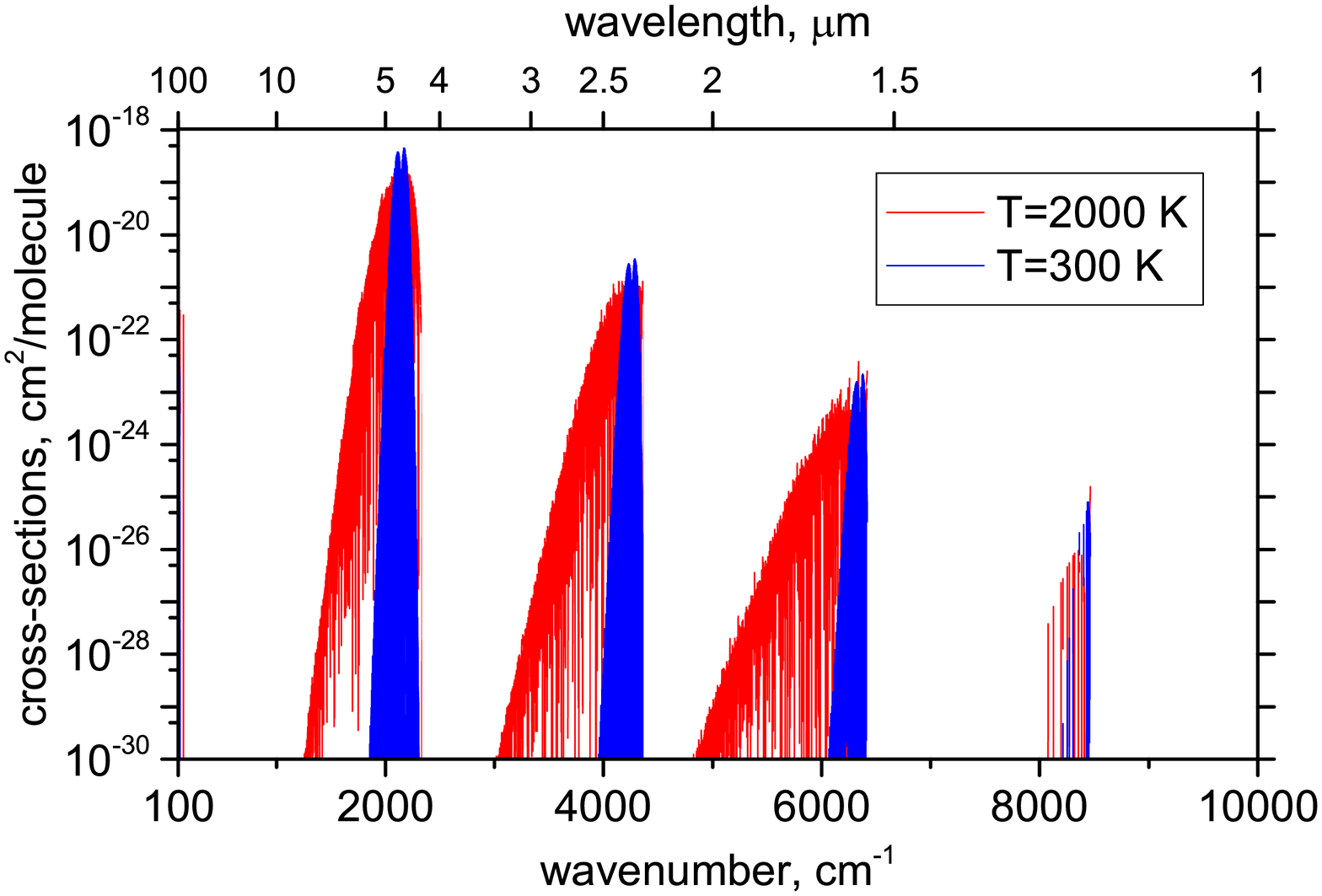}
\caption{Absorption spectrum of CO at $T$ =300~K and 2000~K.}
\label{f:CO}
\end{center}
\end{figure}

{\bf CO:} is the most important diatomic species in a whole range
of hot atmospheres ranging from warm exoplanets to cool stars from
a spectroscopic perspective. Li \ea\  \cite{15LiGoRo.CO}
recently produced comprehensive line lists for the nine main
isotopologues of CO. Figure~\ref{f:CO} illustrates the
absorption spectrum of the main isotopologue, $^{16}$C$^{12}$O.

\begin{figure}
\begin{center}
\includegraphics[width=.86\textwidth]{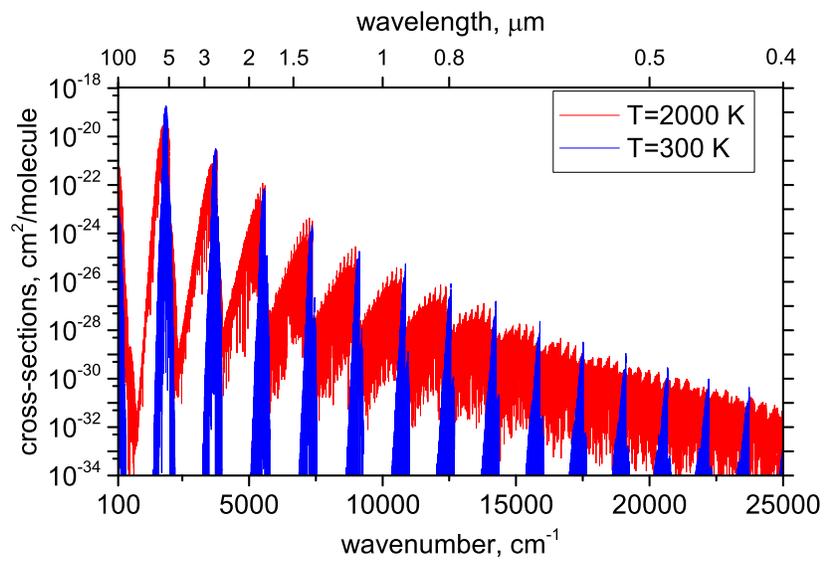}
\caption{Absorption spectrum of NO at $T$ =300~K and 2000~K simulated
using the ExoMol line list \cite{jt686}.}
\label{f:NO}
\end{center}
\end{figure}

{\bf NO:} a new comprehensive line list  for nitric oxide has recently
been released by Wong \ea\  \cite{jt686}, see Fig.~\ref{f:NO}.

{\bf SiO:} Figure~\ref{f:SiO} illustrates the absorption spectrum of SiO molecule.
SiO is well known in sunspots \cite{95CaKlDu.SiO} and is thought likely
to be an important constituent of the atmosphere of hot rocky super-Earths.
An IR line list for SiO available from ExoMol~\cite{jt563} and a less accurate UV line list is provided by Kurucz \cite{11Kurucz.db}.

\begin{figure}
\begin{center}
\includegraphics[width=.86\textwidth]{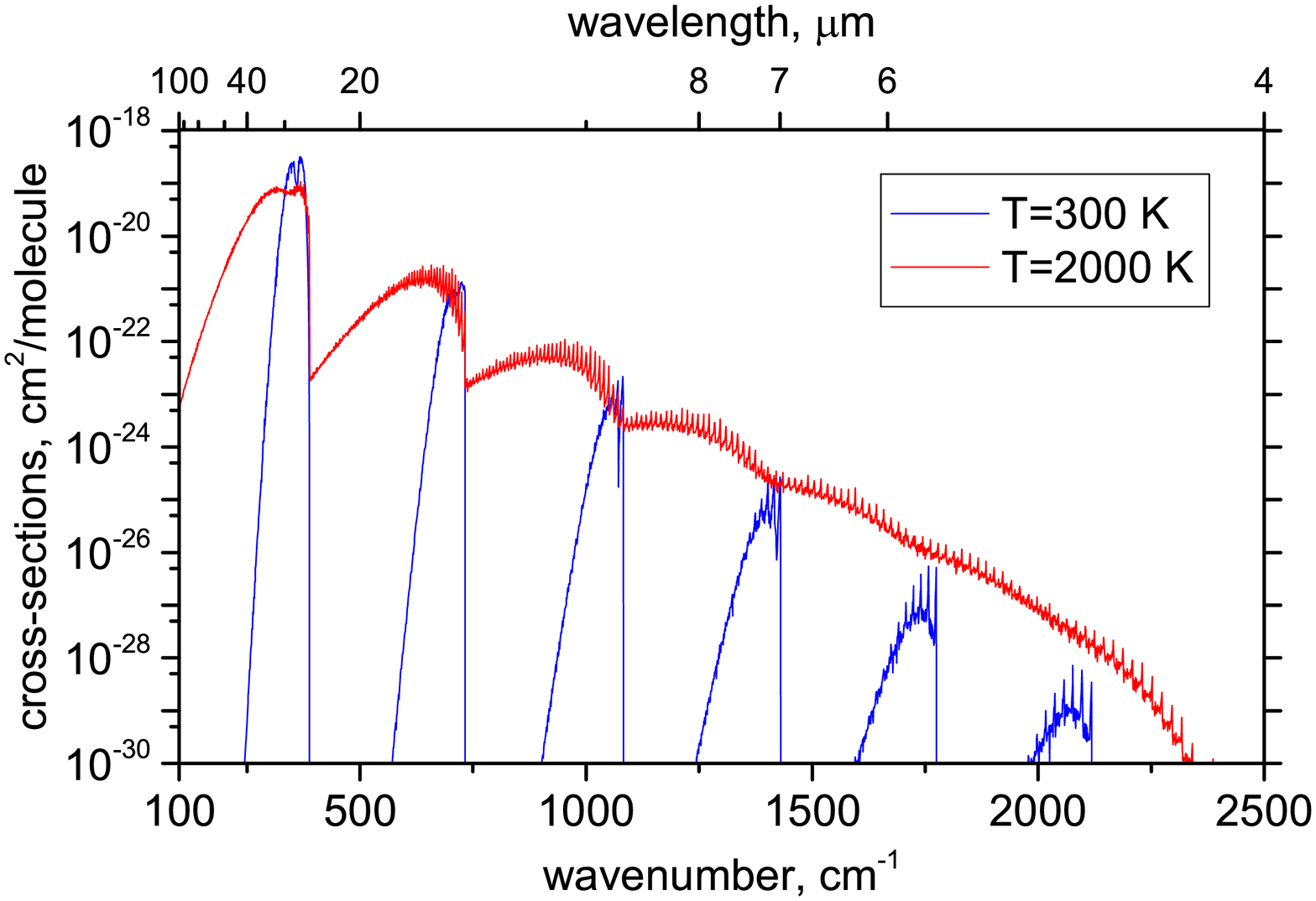}
\includegraphics[width=.86\textwidth]{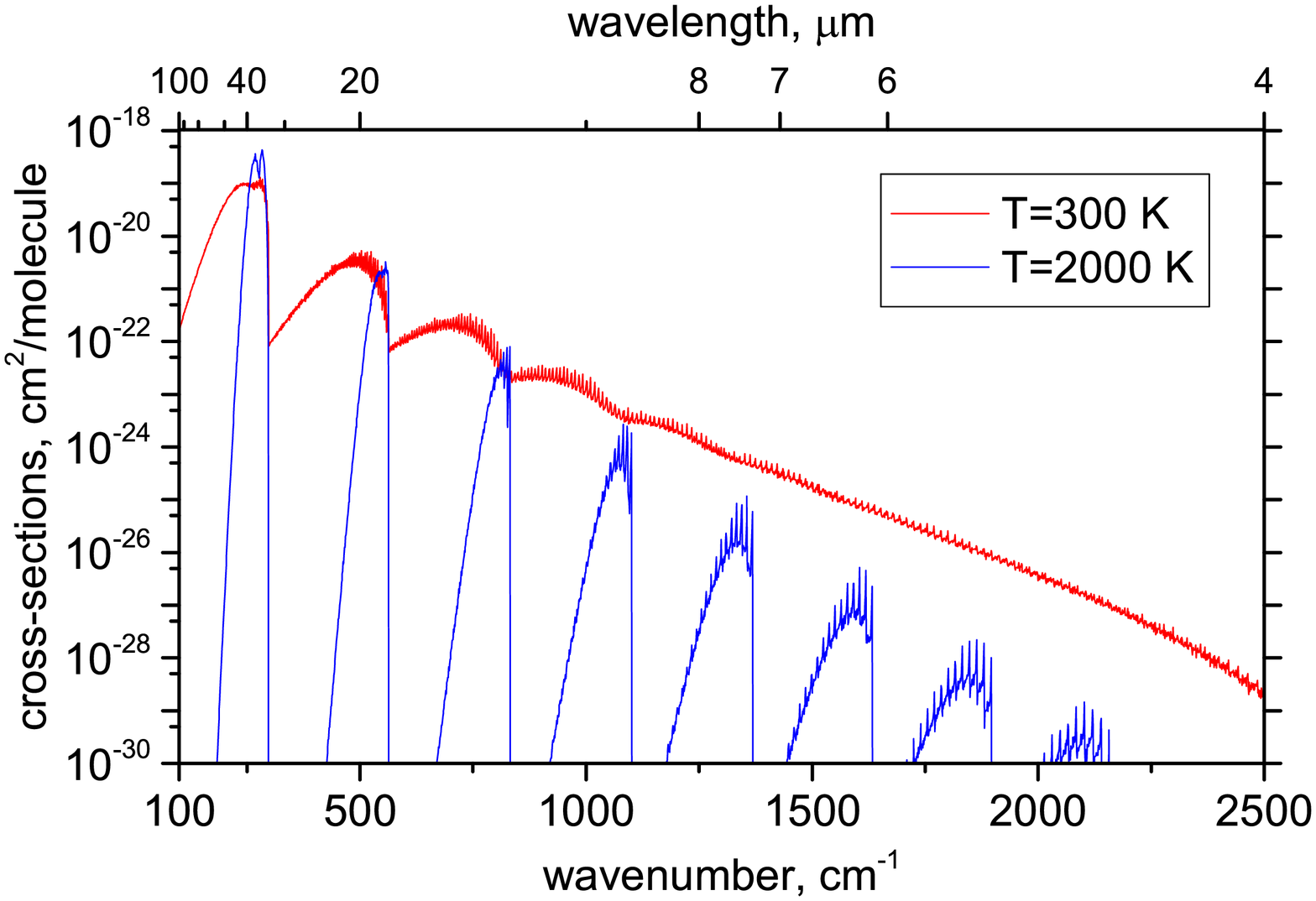}
\caption{Infrared absorption spectra of NaCl (upper) and KCl (lower)
 at $T$ =300~K and 2000~K simulated
using the ExoMol line list \cite{jt583}.}
\label{f:NaCl}
\end{center}
\end{figure}

\begin{figure}
\begin{center}
\includegraphics[width=.86\textwidth]{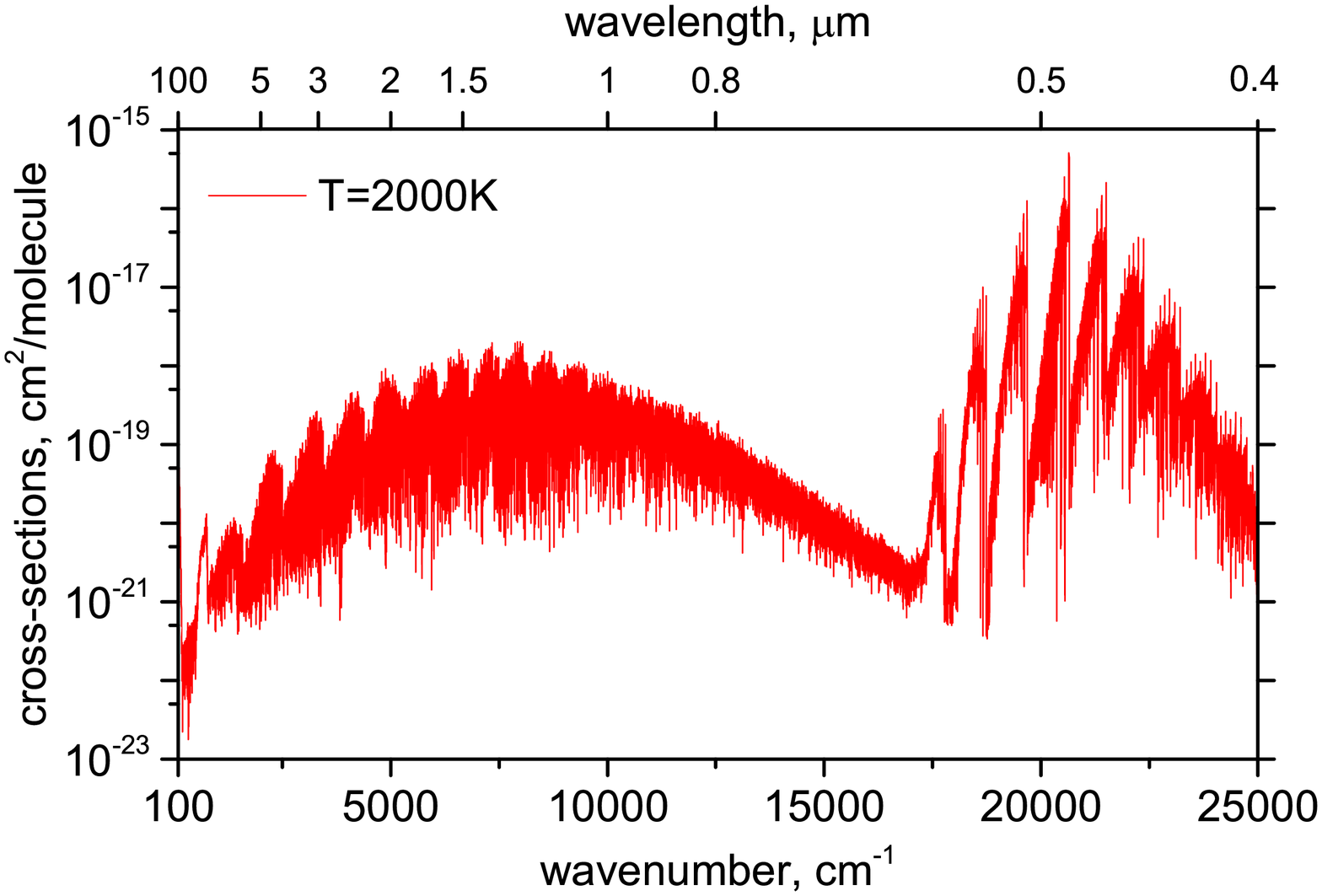}
\caption{Absorption spectrum of AlO at $T$ = 2000~K simulated
using the ExoMol line list \cite{jt598}.}
\label{f:AlO}
\end{center}
\end{figure}

\begin{figure}
\begin{center}
\includegraphics[width=.86\textwidth]{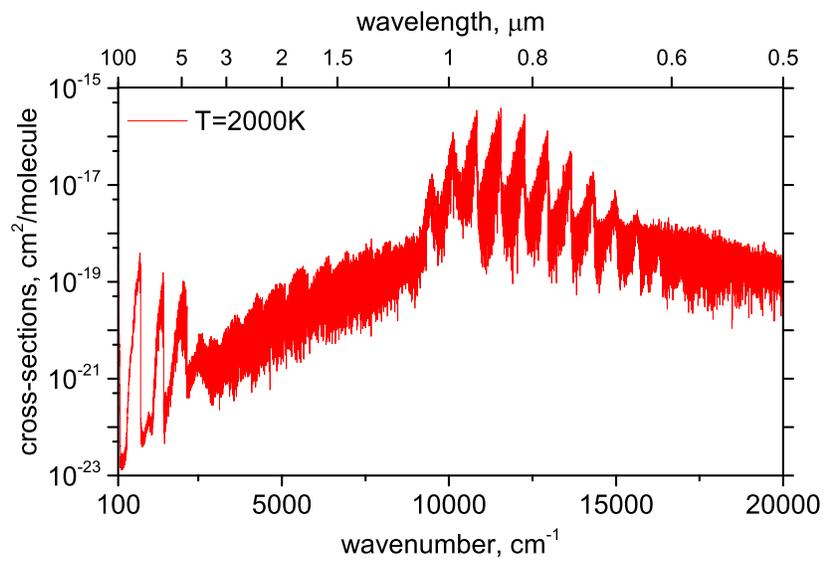}
\caption{Absorption spectrum of CaO at $T$ = 2000~K simulated
using the ExoMol line list \cite{jt618}.}
\label{f:CaO}
\end{center}
\end{figure}

Line lists are available for both NaCl and KCl  \cite{jt583}, see
Fig.~\ref{f:NaCl}. However, these line lists do not consider
electronic transitions, which are likely to be very strong; the line lists
are therefore only useful for simulation
of the spectra of these species at long (infrared) wavelengths.
Figures~\ref{f:AlO} and \ref{f:CaO} illustrate line list for species
whose electronic spectra give prominent features: AlO and CaO respectively.
The spectra are only shown for $T=2000$~K as these species are unlikely
to be found in the gas phase at 300~K.

There are a number of systems which have been identified as likely to
be present in the atmospheres of hot rocky super-Earths for which
there are no available line lists. Indeed for most of these species,
which include NaOH, KOH, SiO$_2$, MgO, PO$_2$, Mg(OH)$_2$, SO, ZnS (see
Table~\ref{species}), there is little accurate spectroscopic data of
any sort. Clearly these systems will be targets of future study.

Probably the most important polyatomic molecule, at least for
exoplanet and cool star research, for which there is still not a
comprehensive hot line list is acetylene (HCCH). Acetylene is a linear
molecule for which variational calculations are possible
\cite{jt346,jt479} and an extensive effective Hamiltonian fit is
available \cite{16AmFaHe.c2h2}. One would therefore expect such
a line list to be provided shortly.



\section{Other considerations}

All the discussion above has concentrated very firmly on line spectra.
However there are a number of issues which need to be considered when
simulating or interpreting exoplanet spectra \cite{15GrHexx}. A discussion of procedures
for this is given in Chapter 5 of the recent book by Heng \cite{17Heng}.
General codes, such as HELIOS \cite{15GrHexx,HELIOS} and our
own ExoCross \cite{ExoCross}, are available for taking appropriate line lists
and creating inputs suitable for radiative transfer codes.

The first issue to be considered is the shape of the individual
spectral lines.  Lines are Doppler broadened with temperature due to
the thermal motion of the molecules and broadened by pressure due to
collisional effects.  While the total absorption by an optically thin
line is conserved as function of temperature and pressure; this is not
true for optically thick lines. For these lines use of an appropriate
line profile can have a dramatic effect \cite{jt521,14AmBaTr.broad}.
The nature of primary transit spectra, where the starlight has a long
pathlength through the limb of the exoplanet atmosphere, is good for
maximizing sensitivity but also maximizes the likelihood of lines
being saturated. This means that it is important to consider line
profiles when constructing line list for exoplanet studies.

While it is straightforward to include the thermal effects via the
Doppler profile; pressure effects in principle depend on the collision
partners and the transition concerned. Furthermore, there has been
comparatively little work on how pressure broadening behaves at high
temperatures \cite{jt584}. Studies are beginning to consider
broadening appropriate to exoplanet atmospheres
\cite{jt544,16HeMaxx,jt669,jt684}. However, thus far these studies have concentrated
almost exclusively on pressure effects in hot Jupiter exoplanets, which means
that molecular hydrogen and helium have been the collision partners
considered. The atmospheres of hot rocky super-Earths are likely to be
heavy meaning that pressure broadening will be important. Clearly
there is work to be done developing appropriate pressure-broadening
parameters for the atmospheres of these planets. We note, however, that
line broadening parameters appropriate for studies of the atmosphere of
Venus are starting to become available, largely on the basis of
theory \cite{00SaMeSe.H2Opb,11GaLaLa.H2Opb,14LaVoNa.h2opb,16GaFaRe.H2Opb}.

Besides broadening, it is also necessary to consider collision induced
absorption in regions where there are no spectral lines. On Earth it is
know that the so-called water continuum majors an important contribution
to atmospheric absorption \cite{12ShPtRa}. Similarly collision induced
absorption (CIA) in by H$_2$ is well known to be important hydrogen
atmospheres \cite{11AbFrLi.CIA}. CIA has also been detected involving K--H$_2$
collisions \cite{15MoMaFo.dwarfs}. What  CIA processes are important
in lava planets is at present uncertain.

Finally it is well-known that the spectra of many (hot Jupiter) exoplanets are
devoid of significant features, at least in the NIR \cite{16SiFoNi,jt699}.
It is thought that this is due to some mixture of clouds and aerosols, often described
as hazes. Such features are likely to also form in the atmospheres of rocky
exoplanets. It remains unclear precisely what effect these will have on the
resulting observable spectra of the planet.

\section{Conclusions}

To conclude, the atmospheres of hot super-Earths are likely to be
spectroscopically very different
those of other types of exoplanets such as cold super-Earth or gas
giant due to both the elevated temperatures and the  different
atmospheric constituents. This means that a range of other species, apart from
the usual H$_2$O,
CH$_4$, CO$_2$ and CO, must be also taken into consideration. A particularly
interesting molecule that is likely to feature  in atmospheric retrievals is SO$_2$. Detection
of SO$_2$ could be used to differentiate super-Venus exoplanets from the broad class of
super-Earths. A comprehensive line list for SO$_2$ is already
available \cite{jt635}.
SiO, on other hand, is a signature of a rocky object with potentially detectable
IR and UV spectral features.  Another interesting species is ZnS, which can be
used to differentiate clouds  and hazes. At present there is no comprehensive line list
for ZnS to inform this procedure.

Models of hot super-Earths suggest that these exoplanets appear to resemble many properties of the early Earth. 
An extensive literature exists on the subject of the early Earth, which can be used as a basis for accurate
prediction of the properties of the hot rocky exoplanets.
Super-Earths also provide  a potential testbed for atmospheric models of the early Earth which, of course, are not
amenable to direct observational tests.
Post-impact planets
may also be also very similar in chemistry and spectroscopy.

From different studies of the chemistry and spectroscopy of hot super-Earth we
have identified a set of molecules suggested either as potential trace species
or sources of opacities for these objects. The line list for a significant number of these
species are either missing or incomplete. Our plan is systematically  create line lists
for these key missing molecules and include into the ExoMol database.

\section*{Acknowledgments}

We thank Giovanna Tinetti, Ingo Waldmann and the members of the ExoMol
team for many fruitful discussion, and Laura Schaefer for
providing a figure. The ExoMol project was supported
by the ERC under the Advanced Investigator Project 267219. This work
was supported by the UK Science and Technology Research Council (STFC)
No. ST/M001334/1. This work made extensive use of the DiRAC@Darwin and
DiRAC@COSMOS HPC clusters.  DiRAC is the UK HPC facility for particle
physics, astrophysics and cosmology which is supported by STFC and
BIS. the support of the COST action MOLIM No. CM1405.  This research
has made use of the NASA Exoplanet Archive, which is operated by the
California Institute of Technology, under contract with the National
Aeronautics and Space Administration under the Exoplanet Exploration
Program.

\bibliographystyle{elsarticle-num}



\end{document}